**Abundances of Deuterium, Oxygen, and Nitrogen in the Local Interstellar Medium: Overview of First Results from the *Far Ultraviolet Spectroscopic Explorer* Mission**

H. W. Moos[1], K. R. Sembach[1,2], A. Vidal-Madjar[3], D.G. York[4], S. D. Friedman[1], G. Hébrard[3], J. W. Kruk[1], N. Lehner[1], M. Lemoine[3], G. Sonneborn[5], B.E. Wood[6], T. B. Ake[1,7], M. André[1,3], W.P. Blair[1], P. Chayer[1,8], C. Gry[9,10], A.K. Dupree[11], R. Ferlet[3], P. D. Feldman[1], J. C. Green[12], J.C. Howk[1], J. B. Hutchings[13], E. B. Jenkins[14], J.L. Linsky[6], E. M. Murphy[15], W.R. Oegerle[5], C. Oliveira[1], K. Roth[16], D. J. Sahnow[1], B.D. Savage[17], J.M. Shull[6,12], T. M. Tripp[14], E. J. Weiler[18], B. Y. Welsh[19], E. Wilkinson[12], B. E. Woodgate[5]

[1]Department of Physics and Astronomy, The Johns Hopkins University, Baltimore, MD 21218; hwm@jhu.edu

[2]Present address: Space Telescope Science Institute, 3700 San Martin Drive, Baltimore, MD 21218

[3]Institut d'Astrophysique de Paris, 98bis Blvd. Arago, 75014 Paris, France

[4]Department of Astronomy and Astrophysics and the Enrico Fermi Institute, University of Chicago, 5640 South Ellis Avenue, Chicago, IL 60637

[5]Laboratory for Astronomy and Solar Physics, Code 681, NASA, Goddard Space Flight Center, Greenbelt, MD 20771

[6]JILA, University of Colorado and NIST, Boulder, CO 80309-0440

[7]Computer Science Corporation, Lanham, MD 20706

[8]Primary Affiliation: Department of Physics and Astronomy, University of Victoria, P.O. Box 3055, Victoria, BC V8W 3P6, Canada

[9]ISO Data Center, ESA Astrophysics Division, P.O. Box 50727, 28080 Madrid, Spain

[10]Laboratoire d'Astronomie Spatiale, BP 8, 13376 Marseille Cedex 12, France

[11]Center for Astrophysics, 60 Garden Street, Cambridge, MA 02138

[12]Center for Astrophysics and Space Astronomy, Department of Astrophysical and Planetary Sciences, University of Colorado, Campus Box 389, Boulder, CO 80309

[13]Herzberg Institute of Astrophysics, National Research Council of Canada, 5071 West Saanich Road, Victoria, BC V8X 4M6, Canada

[14]Princeton University Observatory, Princeton, NJ 08544

[15]Department of Astronomy, University of Virginia, P.O. Box 3818, Charlottesville, VA 22903

[16]Gemini Observatory, Northern Operations Center, 670 N. A'ohoku Pl., Hilo, HI 96720

[17]Department of Astronomy, University of Wisconsin, 475 North Charter Street, Madison, WI 53706

[18]Code S, NASA Headquarters, Washington, DC 20546-0001

[19]Space Sciences Laboratory, University of California, Berkeley, Berkeley, CA 94720

**ABSTRACT**

Observations obtained with the *Far Ultraviolet Spectroscopic Explorer (FUSE)* have been used to determine the column densities of D I, O I, and N I along seven sight lines that probe the local interstellar medium (LISM) at distances from 37 pc to 179 pc. Five of the sight lines are within the Local Bubble and two penetrate the surrounding H I wall. Reliable values of N(H I) were determined for five of the sight lines from *HST* data, *IUE* data, and published *EUVE* measurements. The weighted mean of D I/H I for these five sight lines is $(1.52\pm0.08) \times 10^{-5}$ ($1\sigma$ uncertainty in the mean). It is likely that the D I/H I ratio in the Local Bubble has a single value. The D I/O I ratio for the five sight lines within the Local Bubble is $(3.76\pm0.20) \times 10^{-2}$. It is likely that O I column densities can serve as a proxy for H I in the Local Bubble. The weighted mean for O I/H I for the seven *FUSE* sight lines is $(3.03\pm0.21) \times 10^{-4}$, comparable to the weighted mean $(3.43\pm0.15) \times 10^{-4}$ reported for 13 sight lines probing larger distances and higher column densities (Meyer et al.1998, Meyer 2001). The *FUSE* weighted mean of N I/ H I for five sight lines is half that reported by Meyer et al. (1997) for seven sight lines with larger distances and higher column densities. This result combined with the variability of O I/ N I (six sight lines) indicates that at the low column densities found in the LISM, nitrogen ionization balance is important. Thus, unlike O I, N I cannot be used as a proxy for H I or as a metallicity indicator in the LISM.

*Subject Headings:* cosmology: observations— ISM: abundances— ISM: evolution — Galaxy:abundances—Ultraviolet:ISM

## 1. INTRODUCTION

The measurement of the ratio of deuterium to hydrogen is of particular importance because it provides a constraint on the primordial ratio of D/H, an indicator of the cosmic baryon density (Schramm & Turner 1998; Burles 2000). For three QSO sight lines through the inter-galactic medium (IGM) O'Meara et al. (2001) report an average of D I/H I = $3.0 \pm 0.4 \times 10^{-5}$ and $\Omega_b h^2 = 0.0205 \pm 0.0018$. (All uncertainties in this paper are $1\sigma$ unless otherwise noted.) Recent cosmic microwave background measurements by the BOOMERANG experiment (Netterfield et al. 2001; de Bernardis et al. 2001) and the DASI experiment (Halverson et al. 2001; Pryke et al. 2001) have been used to determine the baryon physical density $\Omega_b h^2 = 0.022 \, {}^{+\,0.004}_{-\,0.003}$ ($H_0 \equiv$ 100km $^{-5}$ Mpc$^{-1}$). Stompor et al. (2001) report a slightly higher value for the MAXIMA-1 experiment, $\Omega_b h^2 \cong 0.033 \pm0.013$ ($2\sigma$), but state that the data is consistent with the lower value. It is reassuring that two very different experimental approaches, the determination of D I/H I in the IGM and measurements of the cosmic background radiation, give such similar answers.

However, uncertainty remains because the IGM results show a larger dispersion than would be expected on the basis of the estimated experimental errors (see O'Meara et al. 2001; Pettini & Bowen 2001). The causes of this large dispersion are uncertain, but may include unknown systematic errors and true cosmic scatter. A recent measurement of D I/H I in an IGM absorber along the sight line toward QSO HS 0105+1619 (O'Meara et al. 2001) samples gas with a size < 1 kpc and density ~ 0.01 cm$^{-3}$, conditions not too different from those in gas in the halo of the Milky Way. The density is only an order of magnitude below that of the Local Interstellar Cloud (LIC). Even with large differences in physical conditions and the amount of stellar processing of the gas, it is reasonable to expect that transport and chemical mixing processes affecting gas in the Milky Way also operate to some (unknown) extent in IGM clouds. Thus, an important first step in understanding the D I/H I determinations in the IGM is to determine how the D I/H I ratio and its constancy are modified by galactic chemical evolution processes occurring within the Milky Way.

The Far Ultraviolet Spectroscopic Explorer (*FUSE*) mission provides a unique capability for measuring the abundance of D I in the interstellar medium (ISM). The isotope shift of the D I Lyman series transitions with respect to those of H I is –82 km s$^{-1}$. *FUSE* detects light in the wavelength region 905 – 1187 Å with a spectral resolution ~15 km s$^{-1}$ (FWHM), permitting access to all of the strong H I and D I Lyman series transitions except for Lyα (Moos et al. 2000; Sahnow et al. 2000a,b). Except along very low column density sight lines, D I Lyα is blended with the strong radiation-damping wings of H I Lyα. Thus, observations of the higher Lyman transitions with weaker absorption strengths are necessary to measure N(D I) along most ISM sight lines. The far-ultraviolet (far-UV) wavelength region contains multiple transitions of D I with a range of absorption strengths. Measurements of several different transitions can often be used together to obtain an accurate deuterium column density along a line of sight. The use of several lines reduces the effects of instrumental uncertainties, the effects of blends with other ISM lines, and the confusion caused by weak stellar features that may overlap particular D I transitions.

In this paper, we report the first D/H results from the *FUSE* mission. The extensive analyses of the seven LISM sight lines studied to date can be found in the companion papers (Friedman et al. 2002; Hébrard et al. 2002; Kruk et al. 2002; Lehner et al. 2002; Sonneborn et al. 2002; Lemoine et al. 2002; Wood et al. 2002). Following an intensive program to focus and characterize the *FUSE* spectrographs, observations to measure D I, O I, N I, and H$_2$ column densities began in early 2000. For H I, we have relied on measurements by *HST*, *IUE*, and *EUVE* for five sight lines; for WD2211-495 and WD0621-376 only *EUVE* measurements without published uncertainties were available. Because the seven sight lines described in the accompanying papers (see Table 1) probe the LISM, the structure of the absorbing gas is relatively simple. Additional studies of other LISM sight lines are underway, and collectively these studies will serve as a foundation for *FUSE* studies of D/H in the Galactic disk and halo.

Measurements of the D I/H I ratio have been reviewed by several authors (e.g. Schramm & Turner 1998, Lemoine et al. 1999, Tytler et al. 2000). We note here that the diffuse ISM is the simplest environment for studying the D/H ratio in the Milky Way, and the most appropriate for comparison with values of D I/H I measured in high-redshift absorption-line systems. The *Copernicus* satellite provided the first measurements of the D/H ratio in the diffuse ISM (Rogerson & York 1973; York & Rogerson 1976; for a survey of *Copernicus* results, see Vidal-Madjar 1991). Later satellite observatories could access only the strong Lyα transition, and measurements were restricted to the very local ISM (see Linsky 1998; Lemoine et al. 1999).

The *FUSE* mission makes it possible to use the Milky Way as a laboratory to understand how the measured deuterium abundances are affected by a wide variety of physical processes. An important step in understanding the nature of such processes is determining whether there are significant variations in the D I/H I (or D I/O I) ratios on scales of less than ~ 100 pc, the approximate dimension of the Local Bubble. The *Copernicus* D I/H I measurements hinted at variations at larger distances (see, e.g., Vidal-Madjar 1991). Moreover, recent results from the Interstellar Medium Absorption Profile Spectrograph (IMAPS) provide compelling evidence for variability on scales of a few hundred parsecs (Jenkins et al. 1999; Sonneborn et al. 2000). These results along with others are presented in § 5 and discussed in § 6.

In order to examine the possibility of variability in the local ISM, Linsky (1998) reviewed the measurements made with GHRS on the Hubble Space Telescope (*HST*) for 12 stars with ISM features at velocities consistent with the LIC. He reported that for the LIC, D I/H I = $(1.50 \pm 0.10) \times 10^{-5}$. The data for clouds other than LIC were more scattered. He concluded that the large uncertainties for many of the measurements precluded a definitive statement about the variation in D/H for more distant sight lines and that higher quality measurements were necessary to address the issue. Because most of the *HST* determinations of D I/H I relied upon absorption profiles observed against cool star emission lines, there has been a concern that there would be systematic differences between values obtained in this manner versus values obtained using hot star continua. In this paper we will show that the *FUSE* D I/H I and D I/O I measurements suggest that the ratios are quite constant out to ~ 100 pc, the approximate dimension of the Local Bubble, whereas a comparison with *Copernicus* and IMAPS data shows an increase in the dispersion of the measurements at longer distances. In addition, the *FUSE* values obtained using hot stars as background sources, are comparable to the *HST* cool star values.

The next section summarizes the properties of the seven *FUSE* sight lines, and compares the properties of the LISM for the various directions. Section 3 contains a discussion the relevant instrumental properties and the data analysis procedures. Both random and systematic uncertainties have been a major concern of the *FUSE* analyses and are discussed in § 4. Section 5 summarizes the results from the seven accompanying papers. A comparison of the results with other measurements and a discussion of the implications are presented in § 6.

## 2. SUMMARY OF *FUSE* TARGETS

Table 1 lists the seven sight lines observed by *FUSE*. The spectral types range from DA white dwarfs with almost featureless stellar continua (e.g., HZ 43A) to subdwarf O stars with a large number of photospheric spectral features (e.g., Feige 110, BD +28º 4211). White dwarfs generally have low values of projected rotational velocity (v sin i < 15 km s$^{-1}$),(Koester et al. 1998). The sight lines cover a large range in both Galactic longitude (54º to 340º) and Galactic latitude (–59º to +84º).

_______________________

INSERT TABLE 1 HERE
_______________________

The ISM in the vicinity of the Sun has been studied extensively, although it is fair to say that our understanding of its content and structure is still incomplete (see Frisch 1995; Ferlet 1999 for reviews).  The Sun is immersed in the LIC, which has a temperature ~7000 K and density n(H I) ~ 0.1 cm$^{-3}$.  This cloud is the primary contributor to the absorption along the sight line to Capella (l = 162.6º, b = 4.6º, d =12.5 pc) (Linsky et al. 1995), which is often used as a benchmark of D I/H I in the ISM.  Figure 1 is a schematic representation of the local structures in the Milky Way plane near the Sun adapted from Cha et al. (2000).  Approximate locations, sizes, and angles subtended by these structures are given by Linsky et al. (2000) and Lallement (1998).  The seven *FUSE* sight lines projected onto the Galactic plane are also shown.  It is tempting to think of ISM clouds as simple homogenous structures, but they clearly contain small scale structure as demonstrated by Na I absorption-line studies (Meyer & Blades 1996; Watson & Meyer 1996).  Even in the limited region of space shown in Figure 1, there are additional clouds and gas structures (see Ferlet 1999 and references therein).  Redfield & Linsky (2000) examined this issue for the LIC and concluded that the LIC appears to be inhomogeneous, but not by a large factor.  The G191-B2B sight line is near the Capella sight line and also samples the LIC.  The model presented by Redfield & Linsky predicts that, for the sight lines to HZ 43A, WD1634-573, and WD2211-495, the LIC contributes less than 5 x 10$^{16}$ cm$^{-2}$ atoms to the total H I column density.  For Feige 110 and BD +28º 4211, the contributions are higher, but are still expected to be less than 1 x 10$^{17}$ cm$^{-2}$.  The contributions of the G cloud to the column densities of WD1634-573 (b = −7º) and WD2211-495 (b = −53º) sight lines are uncertain.  High spectral resolution measurements by *HST* are necessary to determine if there is evidence for a significant amount of material at the velocity of the G cloud.

___________________________

INSERT FIGURE 1 HERE

___________________________

Surrounding the Sun, and outside the LIC and other nearby clouds is a cavity filled with hot low-density gas referred to as the Local Bubble (Breitschwerdt 1998; Snowden et al. 1998, and references therein).  Denser gas having a lower temperature is also present within the cavity.  A high spectral resolution Ca II λ3933.663 study by Crawford et al. (1997) shows velocity structures in the Local Bubble with temperatures and (by implication) densities comparable to those in the LIC.  The existence of nearby Na I structures (Hobbs 1978; Sfeir et al. 1999; Cha et al. 2000) suggests that the neutral gas has T~100 K and n ~ 1 cm$^{-3}$ or larger, but the low values of Na I/Ca II (Welsh et al. 1991; Bertin et al.1993) indicate that the temperatures and densities are likely to be comparable to the LIC for many of the clouds.

Na I absorption-line studies by Sfeir et al. (1999) show that the boundary of the Local Bubble is delineated by a sharp gradient in the neutral gas column density with increasing radius.  The cavity has a radius between 65 and 250 pc depending on direction and is surrounded by a dense neutral gas "wall".  The distance of this neutral gas deduced from X-ray data is generally ~30% smaller (see Snowden et al. 1998).  Sfeir et al. (1999) suggest that this discrepancy can be removed by a proportional adjustment in the X-ray distance scale with concurrent changes in the plasma parameters of the X-ray emission model.  The wall is quite asymmetric so that sight lines in the 1st Galactic quadrant (0º  l  90º) traverse significant amounts of gas at smaller distances than sight lines in the other quadrants (especially the 3$^{rd}$).

Figure 2 shows the projections of the seven *FUSE* sight lines onto the Galactic plane and two other perpendicular planes compared to the 5 mÅ and 20 mÅ Na I λ5891.59 equivalent width contours derived by Sfeir et al. (1999). Discussing the *FUSE* sight lines in the context of these projections requires several notes of caution. The Na I results are preliminary and improvements can be expected for more detailed studies with higher densities of stars per unit solid angle. The uncertainties in the distances to some of the *FUSE* targets can be significant (see Table 1). Furthermore, the boundaries of the cavity are complex. Sfeir et al. (1999) report only three planar cuts through the Galactic plane and two orthogonal meridian planes perpendicular to the plane. Bulges, extensions, and other small-scale structures could affect intermediate cuts. Still, it is possible to make definitive statements about most of the sight lines.

_______________________

INSERT FIGURE 2 HERE

_______________________

WD1634-573 (b = −7º) and WD0621-376 (b = −21º) are well inside the wall of the Local Bubble. G191-B2B (b = 7º) lies near the 5 mÅ Na I contour; the low value of the H I column density (see § 5) implies little or no penetration into the wall, although part of the measured column could come from the region near the wall. The sight line to BD +28º 4211 (b = −19º) penetrates the wall. This is confirmed by examining the plot for the meridian plane containing the l = 90º −270º axis and the Galactic poles (lowest panel of Figure 2). The same plot shows that Feige 110 (b = −59º) also lies beyond the wall. The large H I column densities observed towards these objects confirms this. WD2211-495 (b = −53º) is located at a distance of 53 pc, approximately 14º from the meridian plane containing the l = 0º−180º axis and the Galactic poles (middle panel of Figure 2). Examining this plot shows that the sight line extends past the 5 mÅ Na I contour, but falls short of the 20 mÅ contour by a factor of two in distance; hence, a moderate H I column density is expected. HZ 43A (b = 84º) is close to the north Galactic pole. The low H I column density measured argues that the star does not lie beyond the wall.

## 3. INSTRUMENTAL PROPERTIES

In this section, we discuss the properties of the *FUSE* data and some of the instrumental considerations required to correctly interpret the observed absorption-line profiles. The design of the *FUSE* instrument provides multiple channel wavelength coverage over the 912 − 1187 Å bandpass. A description of the *FUSE* mission and its performance as of early 2000 is given by Moos et al. (2000) and Sahnow et al. (2000a,b). This section presents additional material that was not covered in those papers.

### 3.1 *Wavelength solution*

The relative wavelength accuracy of *FUSE* data in the most recent version of the calibration pipeline used in this work (version 1.8.7) is ~ ±6 km s$^{-1}$ (1σ) over most of the spectrum, with only a few spectral regions having greater uncertainties. The uncertainty in the absolute wavelength scale is larger, particularly for data obtained through the large (LWRS) spectrograph apertures. Pointing uncertainties and optical misalignments can produce zero point shifts up to $\sim\pm100$ km s$^{-1}$, although $\sim\pm30$ km s$^{-1}$ is more typical.  Thus, the *FUSE* absolute wavelength scale was determined independently by comparison of *FUSE* spectral features with related features measured by other instruments (e.g., *HST*, *IUE*, or ground-based telescopes). The zero point uncertainties in the wavelength calibration for apertures other than the LWRS have been significantly improved in version 2.0 of the calibration pipeline released in the fall of 2001, but this was not available at the time of the work reported here.

### 3.2 *Detectors*

The *FUSE* delay-line microchannel-plate detectors are photon-counting devices.  They have low readout noise, permitting long integrations and very good linearity over the dynamic range of the detectors.  Data are recorded either as spectral images aboard the satellite for count rates above $\sim2500$ s$^{-1}$, or for lower count rates as photon address lists that are time-stamped.  The latter mode permits a higher level of filtering for spurious noise during processing on the ground, and is used whenever on-board memory management considerations allow.  At very high count rates, the detector response becomes non-linear.  Thus, at the present time, it is not possible to study the absorption toward any of the very bright stars studied with *Copernicus*. The *FUSE* Operations Team is exploring new techniques for increasing the dynamic range of the satellite, but these were not available at the time of this work.

### 3.3 *Photometric calibration and noise sources*

The current relative photometric calibration of *FUSE* data is accurate to about 10%. Over the several angstrom intervals used to analyze individual ISM absorption features, the relative flux calibration is much better, with several caveats.  First, detector artifacts and scattered light may affect the local flux estimates in the data (see below).  Second, accurate flux calibration in the vicinity of the Lyman series H I lines is inherently difficult because of H I absorption in both the ISM and in stellar atmospheres.  Third, emissions from the upper atmosphere of the Earth can coincide with the absorptions by interstellar H I, O I, and N I.  The effects of terrestrial airglow on the ISM profiles can be evaluated by comparing data obtained during orbital night with data obtained during both day and night.  Most N I and O I emissions are negligible in the night-only data for these studies.   However, H I airglow (particularly H I Ly$\beta$) is always present.

Scattered light and other spurious signals are potential sources of non-random noise in the *FUSE* data.  Broadband signals can usually be removed by a simple subtraction.  (The effects of the line-spread function (LSF) are discussed below in § 3.4.)  Although wavelength-dependent corrections are necessary for very deep exposures (e.g., Kriss et al. 2001), this was not necessary for the relatively bright targets discussed in the papers summarized here.  In some cases, wavelength-dependent count bursts appear for brief, unpredictable times; the data sets have been carefully screened to remove this possible contamination whenever possible.

Airglow also can produce long-term changes in detector response, including geometric distortion, at the emission wavelengths.  These effects depend on the line intensity, the size of the aperture, the proximity of the line to the end of the microchannel plate, and the exposure history of the plate. Such effects have been noted primarily for LWRS spectra taken in the second half of 2000. The effect was reduced by raising the voltage for both detectors on 24 January 2001 and again on 31 July 2001. Much of the data reported here were obtained before this date. In practice, such effects were noted and treated in a manner similar to fixed-pattern noise artifacts as discussed in the next paragraph.

Fixed-pattern noise associated with the detectors has the potential to distort profile shapes and the measured equivalent widths of spectral features. On occasion, the fixed-pattern noise may result in small wavelength shifts as well as affect the observed flux. There are several ways to determine if such problems exist and to correct for them. The four-channel design of *FUSE* gives simultaneous but independent spectra at almost all wavelengths. Typically there are four independent spectra over the central third of the bandpass (~990-1080 Å) and two at other wavelengths. Also, for each spectrograph, there are three entrance apertures that are physically separated in the *y* direction (i.e., perpendicular to the dispersion). These are the large (30"x30", LWRS), medium (4"x20", MDRS), and small (1.25"x20", HIRS) apertures. Thus, it is possible to observe an object at different *y*-locations on the detector during different observations. In the D I/H I analyses discussed here, the spectra from different channels and through different slits were treated independently rather than simply co-added, allowing detector artifacts or other instrumental signatures to be identified.

Fixed-pattern noise in *FUSE* data can be averaged out by motions of the spectrum in the dispersion direction during the course of an observation. Sometimes this is achieved solely by thermally induced motions of the instrument structure as the satellite moves about its orbit. In addition, programmed motions of the focal plane assemblies that hold the entrance apertures can be used to guarantee that averaging will take place in the final co-added data sets. These motions average out most of the small-scale (6–10 pixel) detector artifacts when the individual exposures are aligned in wavelength and averaged. In such cases, signal-to-noise levels are close to those expected from photon statistics, and values ~100 per resolution element have been achieved on the brightest sources (e.g., Sonneborn et al. 2002).

### 3.4 *Backgrounds and Line Spread Function*

Moos et al. (2000) showed that the signal at the center of a broad, strongly saturated C II absorption feature seen in the spectrum of HD 93129A is small, and hence the broadband scattered light is very low. However, the data reported in the accompanying D I/H I papers show signals at a few percent of the continuum at the centers of many of the Lyman series absorption features. Although they have high optical depths, the H I lines are much narrower than the 71 Å width of the C II line used to set a limit on the broadband scattered light. Preliminary modeling of this residual light in the cores of the lines indicates that it can be described by a line-spread function (LSF) with two Gaussian components. One component represents the core of the line and has a width comparable to the instrumental resolution of ~15 km s$^{-1}$. The other is about twice as broad as the core, and is responsible for redistributing light into the center of optically thick lines if they are narrow enough. A comprehensive study of the *FUSE* LSF is not available at this time. However, tests indicate that the effects of the broad component of the LSF generally are small (it probably contains 20 – 30% of the total LSF area) and a large part of the broad component is under the narrow component of the LSF). Additional information regarding the LSF is provided by Kruk et al. (2002). Both single Gaussian LSFs and dual component LSFs have been used in the analyses summarized here. See the individual papers for detailed descriptions of the adopted LSFs.

## 4. DATA REDUCTION AND POTENTIAL UNCERTAINTIES

Several techniques were used to determine the D I, N I, and O I column densities reported in this paper. These included profile fitting of the observed absorption-line profiles, single-component curves of growth fitted to the measured equivalent widths of the lines, and apparent optical depth analyses. These methods rely upon different sets of assumptions and provide complementary information about the strengths of the absorption lines, the presence of unresolved saturated structure, and the velocity structure of the lines. Descriptions of these methods and references to previous use and applicability can be found in the papers devoted to the analyses of the individual sight lines. Here, we note that the various methods employed to analyze the interstellar lines for each sight line generally yield similar results. Furthermore, the data for each sight line were analyzed by at least two independent research teams. The results are an amalgamation of the individual analyses, and the errors adopted account for both the known systematic uncertainties in the analyses as well as the subjective differences between analyses.

Although all of the strong Lyman transitions except for Lyα fall in the *FUSE* wavelength region (905 – 1187 Å), only about thirteen transitions down to Ly-14 at 915.82 Å are useful for making accurate absorption-line measurements of the D I column density. At shorter wavelengths, the overlap between adjacent H I Lyman lines makes the continuum placement highly uncertain. The flux calibration is also uncertain at these wavelengths for the same reason. The line strengths, fλ, for the D I transitions vary by a factor of 150 between Lyβ and Ly-14. Thus, there is access to optically thin transitions over a wide range of column densities. Typically, up to a half dozen transitions may be useful for constraining the column density for a given sight line, although blends with other species, particularly $H_2$, often reduce that number.

A large effort went into determining the sources and magnitudes of the uncertainties in the *FUSE* column densities. The data reduction took place at six different laboratories with frequent reviews and comparisons of techniques and results. Usually, errors due to photon noise were not the dominant contribution. We refer the reader to the seven accompanying papers for comprehensive discussions of the sources of errors relevant for each particular sight line.

As a check on the validity of the error bars assumed in the individual sight line studies, we constructed a simulated set of *FUSE* data for a mock sight line investigation. This blind simulation was analyzed independently by six team members and the results (and errors) were compared to each other and to the parameters of the simulation. The results of this simulation allowed us to assess the impact of known systematic errors on our quoted uncertainties; in some cases the error estimates were too small, in which case we expanded the allowance for the stated error estimates. Throughout this work, we have adopted uncertainties that we believe are conservative and reliably account for known sources of systematic error.

## 4.1 *Velocity Structure*

For the profile-fitting procedures used, it was necessary to treat partially saturated lines properly by employing an appropriate model for the velocity structure of the absorption. Partial knowledge of the velocity structure of the ISM was available from *HST* for some sight lines. Another approach was to rely only on unsaturated lines; see the discussion of WD2211-495 (Hébrard et al. 2002). An associated issue is the effect of the uncertainties in the shape of the line spread function (§ 3.4).

Very low column density H I absorption with a velocity of $-82$ km s$^{-1}$ relative to the primary H I absorption features may also be a source of potential confusion with the observed D I absorption. Such absorption would be undetectable in any of the metal lines (e.g., C II $\lambda1036$, O I $\lambda1039$) if the H I column density is low. Although this is a concern for extragalactic measurements, it is a much smaller concern for the LISM. Clouds with such high velocities are rare in the LISM. In addition, clouds with H I column densities $\sim 10^{15}$cm$^{-2}$ are transparent to ultraviolet ionizing radiation, and thus they are expected to be highly ionized. If low-column density clouds were prevalent in the LISM, they would be observed at all velocities, and a casual examination of *HST, Copernicus* and *FUSE* spectra would commonly show interlopers at various relative velocities. There are rare exceptions (see e.g., Gry & Jenkins 2001), but generally random H I interlopers are not a significant concern for the LISM studies discussed here. The only exception, discussed in §4.4, is near the H I Ly$\alpha$ line where hot optically-thin clouds displaced by modest velocities from the center of the line can have a disproportionate effect on the profile fit if the total H I column density is $\sim 10^{19}$ cm$^{-2}$ or less.

## 4.2 *Continuum Placement*

Properly estimating the continuum is a major consideration for accurate column density determinations, regardless of the analysis method used. Although the spectra of metal-poor white dwarfs are relatively simple, many white dwarf spectra contain photospheric lines. WD0621 -376 (Lehner et al. 2002) and WD2211-495 (Hébrard et al. 2002) are extremely metal-rich DA white dwarfs that show many photospheric far-ultraviolet lines (Holberg et al. 1998). BD +28º 4211 is a sub-dwarf O star with a complex photospheric spectrum (Sonneborn et al. 2002), and the intrinsic stellar spectrum of Feige 110 is also very complex (Friedman et al. 2002). Stellar models were used to guide the continuum placement. However, in the cases of BD +28º 4211 and Feige 110, this was hindered by the complexity of the metal lines and the poorly known atomic data for some of the species arising in the photospheres of these stars.

## 4.3 *Line Blends*

A source of uncertainty in the D I line strengths is introduced by blending with other spectroscopic features. Interstellar H$_2$ (e.g., Shull et al. 2000) is a common source of blends for several of the sight lines studied (see Sonneborn et al. 2002; Friedman et al. 2002). We made a careful examination of atomic and molecular ISM line lists (e.g., Morton 2001; Abgrall et al. 2000) for potential blends. There may also be blends with narrow stellar features that are not properly accounted for in the models. Fortunately, the availability of multiple D I transitions for determining the column densities significantly reduced the impact of this potential problem.

## 4.4 *Determinations of N(H I)*

Even though a large number of H I transitions are contained in the *FUSE* bandpass, determinations of N(H I) are often difficult. Most of the optimum sight lines for *FUSE* measurements of D I do not have accurate values of the H I column available. For the H I column densities reported here ($\sim 10^{18} - 10^{20}$ cm$^{-2}$), the higher order Lyman series lines are typically optically thick, and hence lie on the flat part of the curve of growth. Accurate determinations of the H I column density require measurements of the radiation-damping wings of the Lyα profiles. We used data from the Goddard High Resolution Spectrograph (GHRS) and the Space Telescope Imaging Spectrograph (STIS) on *HST* to estimate N(H I) for BD +28º 4211, G191-B2B and HZ43A. In addition, HZ 43A is exceptional in that the H I column density is low and the properties of the star permit accurate determinations from *EUVE* data. In this case, the GHRS and *EUVE* determinations were in good agreement and were combined. For the Feige 110 sight line, we used high resolution *IUE* data. For three sight lines (WD0621-376, WD1634-573, and WD2211-495), determinations of N(H I) from the photoelectric continua ($\lambda < 912$ Å) measured by *EUVE* were available. *EUVE* measurements of the H I column density are sensitive to the stellar models, and only one of the three (WD1634-573) has a published uncertainty (Napiwotzki et al. 1996; Jordan et al. 1997). The uncertainties for the two other sight lines are estimated to be ±40% (Wolff et al. 1998; Wolff 2001, private communication). Understanding the average value of D/H and the magnitude of its variation in the LISM depends in large part on future *HST* measurements of the H I Lyα profiles and in some cases, better analysis of the *EUVE* measurements.

There are several subtle sources of uncertainty involved in measuring N(H I) from the damping wings of the Lyα line. Uncertainties in the stellar Lyα profile can be important if either the column density is very low, or if the stellar parameters are uncertain. The presence of absorption from the "hydrogen wall" close to the solar heliosphere caused by $H^+ - H^0$ charge exchange with the ISM can have a noticeable effect on the weakest Lyα absorption profiles (Linsky & Wood 1996). Although the H I column densities of this wall are low ($\sim 10^{15}$ cm$^{-2}$), the velocity shifts and high effective temperatures of the material can affect the determination of the total H I for some sight lines. Similar effects could also be caused by other optically-thin high-temperature gas clouds present along the line of sight (see the discussion by Lemoine et al. (2002) for G191-B2B). Hydrogen walls have been noted for other late-type stars in addition to the Sun (Wood et al. 1996; Dring et al. 1997; Wood & Linsky 1998), but little is known about H I in the vicinity of the white dwarf and sub-dwarf stars considered here. Finally, in the spectra of stars with many photospheric spectral features, the presence of undetected metal lines may lead to small increases in the value of the H I column.

## 5. SUMMARY OF RESULTS FOR THE SEVEN SIGHT LINES

Table 2 summarizes the measured column densities for the seven sight lines listed in Table 1. Because we will compare ratios of column densities among different sight lines, 1σ uncertainties (68% probable) are given, while the papers referenced in Table 2 give 2σ uncertainties (95% probable). $H_2$ column densities were measurable only for Feige 110 and BD +28º 4211. Molecular fractions are sufficiently small, $f(H_2) \sim 10^{-4} - 10^{-5}$, that they could be neglected in the subsequent discussions of the total hydrogen column densities in the neutral gas (i.e., N(H I) ≈ N(H I) + 2N(H$_2$)).

Reliable H I column density measurements are available for only five of the seven *FUSE* sight lines. This highlights the need for accurate measurements of H I column densities by using observations outside the *FUSE* spectral range to obtain reliable values of D I/H I. In some cases, it may be necessary to use a proxy for the average H I column density (e.g., O I) as discussed in § 6.

________________________

INSERT TABLE 2 HERE

________________________

Table 3 lists ratios of D I/H I, O I/H I, D I/O I, N I/H I, D I/N I, and O I/N I. Also included is a measure of the average neutral fraction of the gas along the line of sight given by N I/(N I + N II) when available. As in the case of Table 2, the uncertainties are 1σ estimates. We also present the weighted mean of each ratio, its uncertainty, the fractional standard deviation expressed in percent, the number of degrees of freedom (ν), and the $\chi_\nu^2$ test for the weighted mean. The fractional standard deviation is defined as the square root of the weighted average variance of the data (Bevington & Robinson 1992) divided by the weighted mean and is listed as a measure of the actual deviation of the data about the weighted mean. The use of the weighted mean and its uncertainty assumes: 1) the measurements are drawn from a population with a single value, and 2) the reported measurement uncertainties have accounted for all errors. If the $\chi_\nu^2$ test, which compares the actual deviation from the mean to the estimated uncertainty, is significantly greater than one, either or both of the assumptions are called into question. These issues will be addressed in § 6.

________________________

INSERT TABLE 3 HERE

________________________

Table 4 presents previously published D I/H I ratios along other sight lines for comparison with the *FUSE* measurements. Also included are the mean values of O I/H I, N I/H I, and O I/N I for high column density sight lines that extend beyond the Local Bubble (Meyer et al. 1997, 1998). To obtain the average value of O I/N I, we used the five sight lines common to both studies. We have also listed ratios of D I/O I for δ Ori A (Jenkins et al. 1999; Sonneborn et al. 2000; Meyer et al. 1998) and γ Cas (Ferlet et al.1980). To the best of our knowledge, precise measurements of D I/O do not yet exist for any other sight lines in the Milky Way. The *HST* ratios were selected for uncertainties ~15% or less, the *Copernicus* ratios for ~30% or less. Also listed are the three IMAPS values of D I/H I reported by Jenkins et al. (1999) and Sonneborn et al. (2000). We do not include numerous determinations of D I/H I that either have large error bars or additional prior assumptions about the D I/H I value in one or more components along the sight line. We refer the reader to Ferlet et al. (1996) and Linsky (1998) for discussions of these other sight lines.

________________________

INSERT TABLE 4 HERE

________________________

# 6. DISCUSSION

In this section we discuss the mean column density ratios D I/H I, O I/H I, N I/H I, D I/O I, and O I/N I for the seven sight lines observed with *FUSE* and compare these values with measurements for other sight lines observed previously. Finally, we discuss astrophysical processes that may affect the D I/H I and D I/O I ratios.

## 6.1. *The D I/H I Ratios*
### 6.1.1 *D I/H I In and Near the Local Bubble*

For many years, there has been a controversy over the answer to the question, "Does D/H vary along different sight lines in the Galaxy?" (see York & Rogerson 1976; McCullough 1992; Linsky 1998). It is likely that the answer to this question is "yes". Recent IMAPS measurements show a factor of 3 difference between the sight lines to δ Ori A (Jenkins et al. 1999) and γ²Vel (Sonneborn et al. 2000). (See also Figure 3 and the discussion in the text presented later in this section.) These results provide strong evidence that the ratio does in fact vary. The issue now is not whether D I / H I varies, but rather how the ratio varies, the magnitude of the variations, and the distance scales over which the variations occur.

The *FUSE* results show no evidence for significant variations of D I/H I from the mean value for the gas in the Local Bubble and its neutral wall. The five *FUSE* sight lines with reliable H I column densities in Table 2 span more than two orders of magnitude in H I column density. The two high column sight lines (Feige 110 and BD +28º 4211) almost certainly penetrate the wall of the Local Bubble. The weighted mean of D I/H I for the five *FUSE* sight lines with reliable H I column densities is $(1.52 \pm 0.08) \times 10^{-5}$. The fractional standard deviation is 12%. The $\chi_\nu^2$ ($\nu = 4$) test for the weighted mean yields 1.3 indicating that the deviations from the weighted mean are consistent with the estimated uncertainties. The value for Feige 110 is high compared to the mean, D I/H I = $(2.14 \pm 0.41) \times 10^{-5}$. However, the uncertainty is large and removing the Feige 110 value decreases the weighted mean by less than 2%. Including the two sight lines (WD0621-376 and WD2211-495) for which N(H I) is determined from *EUVE* data with a high uncertainty ($\sim 40$ %) does not change these results appreciably. All of the results appear to be consistent with a single value for the Local Bubble. The mean of the *FUSE* values is consistent with the other determinations of D I/H I for the LIC and the Local Bubble. For the sight line to Capella (d=12.5 pc), Linsky et al. (1995) determined D I/H I = $(1.60^{+0.07}_{-0.10}) \times 10^{-5}$. As discussed in § 1, Linsky (1998) reported a mean value of D I/H I = $(1.50 \pm 0.10) \times 10^{-5}$ for the LIC based on 12 sight lines observed with the GHRS. *Thus, the FUSE studies support the idea that the D I/H I ratio is effectively constant out to a distance of $\sim 100$ pc.*

Clearly, small number statistics limit the strength of this conclusion. The number of *FUSE* sight lines with reliable H I column determinations is limited to five, three within the Local Bubble and two penetrating the wall. Also, it is important to note that the *FUSE* data are not of sufficiently high spectral resolution to precisely assess possible variations in D I/H I between individual clouds along a sight line. The integrated values of D I/H I for each sight line are subject to a modest amount of averaging, since even very short lines of sight may have multiple components (see, for example, Piskunov et al. 1997 and Hébrard et al. 1999). However, the integrated values still provide important information about the global properties of the D I/H I ratio.

### 6.1.2 *Comparison With Measurements at Larger Distances*

In Figure 3 we plot the *FUSE* determinations of D I/H I as a function of distance along with the previously measured values of D I/H I from *Copernicus*, *HST,* and IMAPS listed in Table 4 that we believe are particularly well determined. An examination of Figure 3 shows that the variations in D I/H I appear to increase as a function of distance beyond 100 pc. We note that several of the *Copernicus* D I/H I data points for stars beyond the Local Bubble indicate that the D I/H I ratio is substantially smaller than the average local value of 1.5 x10$^{-5}$. For example, the data points for λ Sco (York 1983) and θ Car (Allen et al. 1992) both imply D I/H I ⊑ 1 x10$^{-5}$. In addition, the low IMAPS result for the δ Ori A sight line is a confirmation of the low value measured by Laurent et al. (1979) with *Copernicus*.

_______________________________

INSERT FIGURE 3 HERE

_______________________________

A weighted mean of all of the points in Figure 3 except for WD0621-376 and WD2211-495, which have very large uncertainties, gives D I/H I = (1.39±0.03) x10$^{-5}$. However, $\chi_\nu^2 = 7.1$ ($\nu = 14$), indicating that the probability of a single valued population is extremely low, and hence a weighted mean of D I/H I is not appropriate. If the largest contributor to the $\chi_\nu^2$ test, δ Ori A with D I/H I = $(0.74 \begin{smallmatrix} +0.12 \\ -0.08 \end{smallmatrix})$ x10$^{-5}$, is removed, the weighted mean increases to (1.45±0.03) x10$^{-5}$ but $\chi_\nu^2 = 5.0$ ($\nu = 13$), still indicating that a single valued population for D I/H I is improbable. θ Car has a very low value of D I/H I = (0.50±0.16) x10$^{-5}$ and is the second largest contributor to the $\chi_\nu^2$ test. If it is removed from the test instead, the weighted mean is (1.43±0.03) x10$^{-5}$, but $\chi_\nu^2 = 5.1$ ($\nu = 13$), again indicating a low probability for a single-valued population. We are forced to conclude that although large variations do not appear at distances of ~ 100 parsec or less, it is likely that there are variations at larger distances. Additional *FUSE* measurements of D I and *HST* measurements of H I may help to better characterize this variation.

### 6.2. *The O I/H I and D I/O I Ratios*

#### 6.2.1 *O I/H I*
Measurements of O I/H I are available for five of the *FUSE* sight lines. For the three sight lines in the Local Bubble, the weighted mean is (3.94±0.35) x10$^{-4}$, the fractional standard deviation is 10%, and $\chi_\nu^2 = 0.9$ ($\nu = 2$). Including the two sight lines for which N(H I) is determined from *EUVE* data with large uncertainties (WD0621-376 and WD2211-495) would not change these results appreciably. However, including the two *FUSE* points outside the Local Bubble, BD +28° 4211 and Feige 110 with log(H I) of 19.9 and 20.2 respectively, changes the weighted mean to $(3.03\pm0.21)$ x$10^{-4}$, the fractional standard deviation to 30%, and $\chi_\nu^2 = 3.9$ ($\nu = 4$). Although the ratio is smaller, the fractional standard deviation and $\chi_\nu^2$ are larger than the values for only the Local Bubble sight lines. The changes are dominated by one sight line, BD +28º 4211. With the BD +28º 4211 sight line removed, O I/H I = $(3.93\pm0.32)$ x$10^{-4}$, the fractional standard deviation is 12% and $\chi_\nu^2 = 0.6$ ($\nu = 3$).

Meyer et al. (1998) measured O I/H I for higher column densities, log(H I) = 20.2 to 21.3 versus 17.9 to 18.9 for the Local Bubble points. They found a mean value of $(3.43\pm0.15)$ x$10^{-4}$ for the ISM and argued that the small variations from sight line to sight line indicated that the ratio was essentially constant in the ISM at distances of a few hundred parsecs. (In accordance with Meyer 2001, we have increased their published ratio to correct for a 7.6% decrease in the O I $\lambda1356$ oscillator strength recommended by Welty et al. (1999).) An evaluation of their data shows a fractional standard deviation of 11% and $\chi_\nu^2 = 0.5$ ($\nu = 12$) in agreement with this argument. Within the uncertainties, the *FUSE* weighted means discussed above are comparable with that of Meyer et al.

Meyer et al. (1998) use observations of a single transition, O I $\lambda1356$, and hence a single oscillator strength to determine N(O I) whereas the *FUSE* studies use several O I transitions of varying line strengths, f$\lambda$. With increasing column depth, different transitions are used to determine the value of the column density. In addition, transitions at different wavelengths appear at different locations on the detectors and may have slightly different systematic errors due to instrument artifacts of the type discussed in § 3.3. Thus, relative errors in the O I oscillator strengths and instrumental errors could increase the dispersion in the derived values of O I/H I from transition to transition. However, since several lines and multiple measurements of the same line at different parts of the detector are usually used in the calculations of N(O I) (see, e.g., Hébrard et al. 2002; Sonneborn et al. 2002), the errors in the resulting column densities are expected to be smaller than the potential dispersion due to instrument artifacts and errors in oscillator strengths.

Observed values of O I/H I and D I/O I could be affected by the incorporation of oxygen into interstellar dust. Oxygen in interstellar dust might exist in the form of various silicates such as pryoxene, (Mg,Fe) $SiO_3$ , or olivine, (Mg, Fe)$_2$ $SiO_4$, or as oxides such as MgO, FeO, $Fe_2O_3$ or $Fe_3O_4$. Oxygen could also exist in grain mantles but not in the form of $H_2O$ since the 3.1 μm ice feature is not observed toward stars with low or moderate values of interstellar reddening with E(B-V) < ∼1.0 (Whittet et al. 2001). For the Local Bubble, O I/H I = $(3.91\pm0.33)$ x$10^{-4}$. The level of oxygen depletion derived for the local diffuse ISM from this number depends on the correct reference value for the total ISM oxygen abundance (gas+dust). Here we adopt the Solar abundance as the reference abundance. The Solar abundance of oxygen has recently been revised. Allende Prieto et al. (2001) have measured O/ H = $(4.90\pm0.59)$ x $10^{-4}$ while Holweger (2001) found O /H = $(5.45\pm1.09)$x$10^{-4}$. If we adopt an average of these two numbers, 5.2 x $10^{-4}$, as the ISM total (gas+dust) reference abundance, it suggests the amount of oxygen relative to hydrogen in the dust is ≈ 1.3 x $10^{-4}$. This estimate is similar to the value expected, ∼(1.5 to1.8) x $10^{-4}$, if the oxygen in the dust is mostly in silicate and oxide grains (Mathis 1996). Our estimate implies that ∼ 25 % of the interstellar oxygen in the Local Bubble could reside in interstellar dust grains. Variations in the amount of dust due to differences in the history and environment of a cloud could affect the measured values of O I/H I and D I/O I at a low level. For sightlines with high values of interstellar reddening larger effects are possible.

### 6.2.2   D I/O I

Because reliable H I measurements are not always available, it is desirable to inquire if the ratio D I/O I can be used instead of D I/H I to search for variability.  Timmes et al. (1997) have argued that D I/O I and D I/N I may provide unique advantages over D I /H I for tracing the evolution of D as a function of metallicity, time, and redshift.  Even though these arguments are aimed at studies of intergalactic gas, a similar approach is attractive for the Milky Way.  The ionization balances of O I and D I are strongly coupled through charge exchange reactions with H I.  Meyer et al. (1998) have shown that it is likely that the sight line to sight line variations in O I/H I are small in the nearby ISM and the *FUSE* results presented here support this for the Local Bubble.  The weighted mean of D I/O I for the five *FUSE* sight lines within the Local Bubble is $(3.76\pm0.20)$ x$10^{-2}$.  The fractional standard deviation is 12 % and $\chi_\nu^2 = 1.1$ ($\nu = 4$) indicating that the variability of the data is consistent with the uncertainties. A *FUSE* survey of D I/O I in the Local Bubble by Hébrard et al. (2001) that includes three additional sight lines also concludes that this ratio is constant.  The constancy of the D I/O I ratio in the Local Bubble within the observational limitations is in agreement with the results for D I/H I presented in § 6.1.  Thus it is likely that, at least in an average sense, O I could serve as a proxy for H I in the Local Bubble, and possibly beyond in the nearby ISM.

Although the D I/O I ratio appears to be constant in the Local Bubble, is it appropriate to use a single mean value to describe the D I/O I values at larger distances?   If one includes the two more distant *FUSE* sight lines (Feige 110 and BD +28° 4211) with the values inside the Local Bubble, D I/O I = $(3.99\pm0.19)$ x$10^{-2}$, the fractional standard deviation is 20 % and $\chi_\nu^2 = 2.6$ ($\nu = 6$).  If the two additional values in Table 4 ($\gamma$ Cas and $\delta$ Ori A) are included, the weighted mean is $(3.63\pm0.16)$ x$10^{-2}$, the fractional standard deviation is 26% and $\chi_\nu^2 = 3.8$ ($\nu = 8$).  The increase in the value of $\chi_\nu^2$ indicates that the variability in the data is larger than would be expected from the reported measurement errors.  Furthermore, the ionization balances of O I and D I are strongly coupled through charge exchange reactions with hydrogen, so the observed scatter is unlikely to be caused by a selection effect involving ionization of one species, but not the other.  Thus, although unknown systematic errors cannot be ruled out, the increases in fractional standard deviation and $\chi_\nu^2$ at larger column density and distance may be due in part to real variations in the value of D I/O I beyond the Local Bubble.  These variations may result from changes in the relative concentrations of D I, similar to the variations of D I/H I with increasing distance displayed in Figure 3.

### 6.2.3 O I/H I versus D I/H I and D I/O I versus O I/H I

Even though the variations in O I/H I are expected to be small, it is important to investigate whether there is any anti-correlation between O I and D I.  Because D is destroyed while O is produced by nucleosynthesis, there could be an anticorrelation between the column densities of the two atoms.  If so, is such an effect observable over the limited metallicity range of the nearby ISM?  Figure 4 displays values of O I/H I compared to D I/H I for the seven *FUSE* sight lines plus two additional points from Table 4.  D I/H I varies over a factor 2.9 from $\delta$ Ori A to Feige 110. As discussed in § 4.4, we have adopted 40% errors on N(H I) for the *FUSE* determinations of D I / H I and O I / H I for the WD0621-376 and WD2211-495 sight lines. Note that errors in the H I column density tend to move points along a diagonal line with positive slope whereas an anti-correlation would move the points along a line with negative slope. Figure 4 displays no evidence for anti-correlation.

---------------------------------

INSERT FIGURE 4 HERE

---------------------------------

The D I/O I ratio should be even more sensitive to anti-correlation effects than the data presented in Figure 4. Figure 5 shows the relationship between D I/O I and O I/H I. An anti-correlation of D I/O I and O I/H I would manifest itself as a trend running from the upper left to lower right in the figure. (Note that systematic errors in the O I columns will produce the same effect.) The data points show no clear trends over the limited metallicity range shown. A "scatter plot" is expected if D I is not anti-correlated with O I as implied by Figure 4, and Figure 5 appears to confirm this expectation. Over large ranges of metallicity, e.g. between the Milky Way ISM and the IGM, a strong anti-correlation between D and O should exist. It is not yet known whether a similar effect exists within the Milky Way ISM. However, Figures 4 and 5 indicate that the abundances of D I and O I are not anti-correlated at an observable level in the nearby ISM.

---------------------------------

INSERT FIGURE 5 HERE

---------------------------------

### 6.3. *The N I/H I, O I/N I and D I/N I Ratios*

Local conditions in the ISM can affect the ionization balance of N I with respect to H I, D I, and O I. In environments where the self-shielding is small and the electron density significant, which is often true for LISM sight lines, the charge exchange coupling of N I with H I is much weaker compared to that of O I with H I. For the three *FUSE* sight lines where estimates of N(N II) are available, the fraction of N in the form of N I is roughly one third (Table 3). Vidal-Madjar et al. (1998) measured a lower than average value of N I/H I along the sight line to G191-B2B, and all of the studies reported here show low values of N I/H I. Jenkins et al. (2000) performed a *FUSE* study of four white dwarfs in the LISM and noted similar effects. The observed deficiencies of N I are most likely due to photoionization by radiation from hot stars and recombination of gas within hot conduction fronts. In the model discussed by Jenkins et al. (2000), the local N I deficiency monotonically decreases to -0.07 dex with increasing N(H I) at N(H I) = $10^{18}$ cm$^{-2}$. The actual deficiency measured for any given sight line will depend on the exact conditions encountered along the path, but the implication is that for low column density clouds in the LISM the ionization balance of N I compared to H I is expected to be more variable than that of O I. As a result, at low column densities N I is probably not a reliable metallicity indicator.

The abundance of N I relative to H I, D I, and O I appears to be lower in the nearby ISM probed by the *FUSE* sight lines compared to directions having larger neutral hydrogen column densities. Meyer et al. (1997) studied seven lines of sight with log N(H) = log[N(H I) + 2N(H$_2$)] ranging from 20.18 to 21.15 and found N I/H I = (7.5±0.4) x10$^{-5}$, near the solar value of (9±2) x10$^{-5}$ (Holweger 2001). The weighted mean reported here for the five *FUSE* sight lines with high quality H I measurements is a factor of 2 less than that of Meyer et al. The average measured N I column appears to be low. The *FUSE* value of O I/N I is twice as large as that calculated from the data of Meyer et al. (1997, 1998), leading to a similar conclusion. York et al.(1983) found a similar deficiency of N I compared to O I in a sample of 53 sight lines observed by *Copernicus*. It is likely that the apparently lower N I abundance along the *FUSE* sight lines is an ionization effect and not a metallicity effect. For D I/N I the *FUSE* weighted mean is 60% larger than that for the three IMAPS sight lines: $\gamma^2$ Vel, $\zeta$ Pup, and $\delta$ Ori A (Sonneborn et al. 1999), which suggests again that ionization of N I is more important in the *FUSE* sight lines with lower total column density.

### 6.4. *Causes of Variation in the Abundance Ratios*

#### 6.4.1 *Implications for Interstellar Mixing*

The results of § 6.1 indicate that it is reasonable to approximate the value of D/H within ~100 pc of the Sun by a single value with a relatively small scatter. The validity of this assumption for other regions of the Galactic disk is less secure from an observational standpoint, as outlined earlier. A simple explanation for why differences in the D/H ratio may be seen in the Galaxy over distances separated by only a few hundred parsecs can be cast in terms of the efficiency of mixing processes within the ISM driven by supernovae (SNe). There are two relevant time scales: the mixing time, and the time between SNe. Ideally, the first of these can be represented by the signal crossing time, or the time it takes for one parcel of gas to be influenced by another parcel of gas. In the simplest case, the second is the inverse of the supernova rate.

Consider a spherical region 100 pc in radius, similar in size to the Local Bubble considered within this study. In a fully ionized ISM at ~ 10$^6$ K, the adiabatic sound velocity $c_S \approx$ 150 km s$^{-1}$. For r = 100 pc, the crossing time is $t_S \approx$ 7 x 10$^5$ yr. This is roughly the crossing time expected for the rarefied regions of the Local Bubble. Other mechanisms are generally slower. The Alfvén velocity is the appropriate signal speed for travel through the cloudy ISM threaded by magnetic fields. The Alfvén velocity is $v_A$ = 17 km s$^{-1}$ for a cloud with a magnetic field of three microgauss and n ≈ 1 cm$^{-3}$; the Alfvén crossing time is $t_A \approx$ 6 x 10$^6$ yr. $v_A$ is proportional to B, so smaller magnetic fields would lead to longer mixing times. The LIC moves with a velocity of 19 km s$^{-1}$ LSR away from the Sco-Cen association. Other nearby clouds have similar velocities (see Lallement 1998). Breitschwerdt (1998), Egger (1998), and Breitschwerdt et al. (2000) have presented a model in which gas parcels with dimensions of a few parsecs are ejected from the region of the Local Bubble wall in the Sco-Cen direction and travel through the Local Bubble at ballistic velocities. Internal magnetic fields prevent the clouds from ablating during transit. For ballistic clouds, the crossing time is $t_B \approx$ 4 x 10$^6$ yr.

Assuming a uniform supernova production rate of 0.02 yr$^{-1}$ (Cappellaro et al. 1993) spread over a galactic disk with a radius of 8.5 kpc, one expects roughly 8.8 x 10$^{-11}$ SNe pc$^{-2}$ yr$^{-1}$.

The typical time between SNe in a 100 pc radius region is 4 x $10^5$ yr, so $t_{SN} \sim t_S$ for this idealized situation of uniform star formation. Of course, some regions form stars more actively than others; most Type II SNne occur in OB associations. Consequently, one expects that regions where $t_{SN} << t_S$ (e.g., spiral arm regions like Orion or Vela) may have considerable inhomogeneity compared to more quiescent regions (like the solar neighborhood) because astrated gas does not have a chance to mix completely before newly processed gas is expelled into the region. Also, the correlated nature of SNe reduces the average frequency of mixing events. Assuming ~ 40 SNe progenitors (OB stars) per association would increase the value of $t_{SN}$ by 40 where $t_{SN}$ now refers to the average time between each burst of star formation and SNe. However, each association has a "sphere of influence" with r ~ $10^2$ pc so that the region for calculating the SNe rate doubles to ~ 200 pc and $t_{SN} \sim 4$ x $10^6$ yr. Thus, the crossing times calculated above indicate that the ISM within ~ 100 pc should be well mixed during this interval. Conversely in the extreme case, if $t_{SN} >> t_S$, the amount of energy injected into the ISM may be insufficient to thoroughly mix the gas.

More thorough discussions of the cycling of heavy metals between stars and the ISM generally conclude that the typical mixing times required to produce a homogeneous ISM are long, t > $10^8$ yr (Avillez 2000; Tenorio-Tagle 2000; Avillez & MacLow 2001). A variety of factors determine the mixing efficiencies, and many of these have not yet been studied observationally in sufficient detail to fully characterize the mixing timescales (e.g., diffusion, magnetic fields, small-scale interstellar structure, the transfer of energy between hot plasmas and cold clouds). We suggest that the D/H results contained herein provide an additional constraint on models seeking to explain the recycling and dispersal of elements in the ISM. Ongoing D I/H I observations with *FUSE* may shed additional light on this subject.

### 6.4.2 *Nucleosynthetic Histories*

The fact that D I/H I appears to have a single value for gas within 100 pc of the Sun, whereas at larger distances there appear to be variations in the ratio, may indicate that the different regions have had very different nucleosynthetic histories. This result is even more puzzling when one considers that O I/H I does not vary as much as D I/H I in the near ISM (see § 6.2). In addition, over the limited metallicity range available in the nearby ISM, the abundance of D I does not vary inversely with the abundance of O I. This contradicts the simple idea that metal production and D destruction occur together. Such a picture is probably valid over large (factor of 10-100) differences in metallicity, but there may exist nucleosynthetic mechanisms in the Milky Way that are more subtle. For example, the binding energy of D I is so low (~2.2 MeV) that it is destroyed at temperatures much lower than those required for the production of O in massive stars. Thus, if D is destroyed in the outer layers of stars and is ejected before significant O production occurs, the average abundance of D could decrease without increasing the average metallicity. Such a process could be operative for early-type stars that shed their outer envelopes in massive stellar winds. The relative abundance of D, O, and H in the wind will vary with time (and final location), but the detailed spatial variation would depend on the stellar population in a given region. In addition, the average number of massive stars is small, so that the degree of astration and metal production will fluctuate from location to location.

It is also possible that there may be sources of D that could provide variations from place to place in the galaxy. Mullan & Linsky (1999) suggest that contamination of the ISM by D-enriched material ejected from stellar flares is such a source. Other sources, such as spallation of heavier elements (Epstein et al. 1976), have also been proposed although all have some drawbacks. Lemoine et al. (1999) summarize the plausibility that these and other processes contribute to the observed D abundances in the LISM.

## 7. CONCLUSIONS

The unique access of *FUSE* to the far-UV spectral region has been used to measure column densities of D I, O I, and N I for the seven sight lines listed in Table 1. The authors of the accompanying seven papers describe the care taken to assess the accuracy of the uncertainties in these measurements. Collectively, the sight lines probe the LIC, the G cloud, other clouds in the Local Bubble, and the wall of the Local Bubble. Although the sight lines have a range of a factor of ~5 in distance, the H I column density has a range of ~160, indicating the wide variety of environments sampled.

H I columns and uncertainties were obtained from *HST* data (two sight lines), *IUE* data (one sight line), one *EUVE* value, and in one case a combination of *HST* and *EUVE* results. For these five sight lines, the weighted mean of D I/H I = $(1.52\pm0.08) \times 10^{-5}$. This mean is similar to previous values obtained for the LIC. A $\chi_\nu^2$ test of the weighted mean and other tests show that it is likely that D I/H I can be represented by this single value out to ~ 100 pc, the approximate dimension of the Local Bubble. However, an examination of *Copernicus* and IMAPS data at larger distances shows an increased dispersion.

The mean value of O I/H I is similar to that reported by Meyer et al. (1998) for longer sight lines. For D I/O I within the Local Bubble, the dispersion is consistent with the estimated uncertainties, in agreement with the results for D I/H I. It is likely that O I can be used as a proxy for H I in the Local Bubble and possibly in the nearby ISM. When data taken by *FUSE* and other missions for longer sight lines is included, the dispersion grows. Although unknown systematic errors cannot be ruled out, this may be due in part to variability in D I, similar to that observed for D I/H I.

These results are subject to small number statistics. *FUSE* is making additional measurements of D I and O I, but *HST* measurements of the H I column density and the gas velocity structure for many of these sight lines are critical for further progress.

An estimate of mixing times for the ISM shows that mixing of a region the size of the Local Bubble is consistent with the frequency of SNe. In regions that form stars more actively than others (e.g., spiral arm regions such as Orion or Vela), it may not be possible to mix the gas between bursts of stellar formation and they may show considerable inhomogeneity compared to the solar neighborhood. Although the nucleosynthetic destruction of D in stars is expected to be accompanied by an increase in O, comparisons of O I/H I versus D I/H I and D I/O I versus O I/H I do not show any evidence for an anti-correlation between D I and O I over the limited range of metallicity studied. This may require new stellar processing mechanisms such as the destruction of the weakly bound D nucleus without accompanying O I production in the outer layers of a hot star followed by ejection of the depleted material in massive stellar winds. The possibility of local sources of D also needs to be considered.

There also is a cosmic dispersion in the values of D I/H I measured at large distances (Pettini & Bowen 2001). As a consequence, determinations of baryon densities from D I/H I must be based on careful statistical studies. However, searches for suitable systems to study have a very low yield, and there is a danger of subtle selection biases. Thus, it is important to understand the cause(s) of the observed dispersion in D I/H I. Typical gas volume densities in the damped Ly$\alpha$ systems are close to those in the outer parts of the Milky Way (O'Meara et al. 2001), and it is likely that future studies leading to a better understanding of the still unknown causes of the dispersion in the ISM will illuminate those of the IGM.

This work is based on data obtained for the Guaranteed Time Team by the NASA-CNES-CSA *FUSE* mission operated by the Johns Hopkins University. Financial support to U. S. participants has been provided by NASA contract NAS5-32985. French participants have been supported by CNES. This study of the LISM required the efforts of a number of people and we acknowledge their efforts in assembling this study. We also acknowledge the effort of *FUSE* operations scientists and staff in planning the observations and preparing the data for analysis. D. Sfeir kindly provided the Na I profiles used in Figure 2. We thank Sharon Tiebert-Maddox for assistance with the preparation of the manuscript and figures.

REFERENCES

Abgrall, H., Roueff, E., & Drira, I. 2000, A&AS, 141, 297

Allen, M. M., Jenkins, E. B., & Snow, T. P. 1992, ApJS, 83, 261

Allende Prieto, C., Lambert, D. L., & Asplund, M. 2001, ApJ, 556, L63

Avillez, M.A. 2000, MNRAS, 315, 479

Avillez, M. A., & Mac Low, M.-M. 2001, Astronomische Gesellschaft Abstract Series, Vol. 18

Bertin, P., Lallement, R., Ferlet, R., & Vidal-Madjar, A. 1993, A&A, 278, 549

Bevington, P. R. & Robinson, D. K. 1992, Data Reduction and Error Analysis for the Physical Sciences, New York: McGraw-Hill, 2nd ed.

Breitschwerdt, D. 1998, Berlin Springer Verlag Lecture Notes in Physics, 506, 5

Breitschwerdt, D., Freyberg, M. J., & Egger, R. 2000, A&A, 361, 303

Burles, S. 2000, Nuclear Physics A, 663, 861

Cappellaro, E., Turatto, M., Benetti, S., Tsvetkov, D. Y., Bartunov, O. S. & Makarova, L. N. 1993, A&A, 273, 383

Cha, A. N., Sahu, M. S., Moos, H. W., & Blaauw, A. 2000, ApJS, 129, 281

Crawford, I. A., Craig, N., & Welsh, B. Y. 1997, A&A, 317, 889

de Bernardis, P., et al. 2001, ApJ, submitted, astro-ph/0105296

Dring, A. R., Linsky, J., Murthy, J., Henry, R. C., Moos, W., Vidal-Madjar, A., Audouze, J., & Landsman, W. 1997, ApJ, 488, 760

Egger, R. 1998, Berlin Springer Verlag Lecture Notes in Physics, v.506, 287

Epstein, R. I., Lattimer, J. M., & Schramm, D. N. 1976, Nature, 263, 198

Ferlet, R. 1999, A&ARV, 9, 153

Ferlet, R., Vidal-Madjar, A., Laurent, C., & York, D.G. 1980, ApJ, 242, 576

Ferlet, R. et al. 1996, in Science with the *HST*-II, ed. B. Benvenuti et al. (Space Telescope Science Institute: Baltimore), p. 450

Friedman, S. D., et al. 2002, ApJS, submitted

Frisch, P. C. 1995, Sp. Sci. Rev., 72, 499

Gry, C., & Jenkins, E. B. 2001, A&A, 367, 617

Halverson, N. W., Leitch, E. M., Pryke, C., Kovac, J., Carlstrom, J. E., Holzapfel, W. L., & Dragovan, M. 2001, ApJ, submitted, astro-ph/0104489

Hébrard, G., et al. 2002, ApJS, submitted

Hébrard, G., et al. 2001, XVIIth IAP Colloquium, *Gaseous matter in galaxies and intergalactic space*, 19-23 June 2001, Paris, Edited by R. Ferlet et al., to be published

Hébrard, G., Mallouris, C., Ferlet, R., Koester, D., Lemoine, M., Vidal-Madjar, A., & York, D.G. 1999, A&A, 350, 643

Hobbs, L. M. 1978, ApJ, 222, 491

Holberg, J. B., Barstow, M. A., & Sion, E. M. 1998, ApJS, 119, 207

Holweger, H. 2001, Astro-ph. 0107426

Jenkins, E. B., et al. 2000, ApJ, 538, L81

Jenkins, E. B., Tripp, T. M., Woʟhiak, P., A., Sofia, U. J., & Sonneborn, G. 1999, ApJ, 520, 182

Jordan, S., Napiwotzki, R., Koester, D., & Rauch, T. 1997, A&A, 318, 461

Koester, D., Dreizler, S., Weidemann, V., & Alland, N. F. 1998, A&A, 338, 612

Kriss, G. A., et al. 2001, Science, 293, 1112

Kruk, J. W., et al. 2002, ApJS, submitted

Lallement, R. 1998, Berlin Springer Verlag Lecture Notes in Physics, 506, 19

Lehner, N., et al. 2002, ApJS, submitted

Lemoine, M., et al. 2002, ApJS, submitted

Lemoine, M., et al. 1999, New Astronomy, 4, 231

Linsky, J. L. 1998, Sp. Sci. Rev., 84, 285

Linsky, J. L., Diplas, A., Wood, B. E., Brown, A., Ayres, T. R., & Savage, B. D. 1995, ApJ, 451, 335

Linsky, J. L., Redfield, S., Wood, B. E., & Piskunov, N. 2000, ApJ, 528, 756

Linsky, J. L., & Wood, B. E. 1996, ApJ, 463, 254

Mathis, J. S. 1996, ApJ, 472, 643

Meyer, D. M. 2001, XVIIth IAP Colloquium, Gaseous Matter in the galaxies and intergalactic space, 19-23 June 2001, Edited by R. Ferlet et al., to be published

Meyer, D. M., & Blades, J. C. 1996, ApJ, 464, L179

Meyer, D. M., Cardelli, J. A., & Sofia, U. J. 1997, ApJ, 490, L103

Meyer, D. M., Jura, M., & Cardelli, J. A. 1998, ApJ, 493, 222

Moos, H. W. et al. 2000, ApJ, 538, L1

Morton, D.   2001, in preparation

Mullan, D. J., & Linsky, J. L. 1999, ApJ, 511, 502

Napiwotzki, R., Jordan, S., Bowyer, S., Hurwitz, M., Koester, D., Rauch, T., & Weidemann, V. 1996, IAU Colloq. 152: Astrophysics in the Extreme Ultraviolet, 241

Netterfield, C. B., et al. 2001, ApJ, submitted, astro-ph/0104460

O'Meara, J. M., Tytler, D., Kirkman, D., Suzki, N., Prochaska, J. X., Lubin, D., & Wolfe, A. M. 2001, ApJ, 552, 718

Pettini, M., & Bowen, D. V., 2001 astro-ph/0104474

Piskunov, N., Wood, B. E., Linsky, J. L., Dempsey, R. C., & Ayres, T. R. 1997, ApJ, 474, 315

Pryke, C., Halverson, N. W., Leitch, E. M., Kovac, J., Carlstrom, J. E., Holzapfel, W. L., & Dragovan, M. 2001, ApJ, submitted, astro-ph/0104490

Redfield, S., & Linsky, J. L. 2000, ApJ, 534, 825

Rogerson, J. B., & York, D. G. 1973, ApJ, 186, L95

Sahnow, D. J., et al. 2000a, ApJ, 538, L7

Sahnow, D. J., et al. 2000b, Proc. SPIE, 4013, 334

Schramm, D. N., & Turner, M. S. 1998, Reviews of Modern Physics, 70, 303

Sfeir, D. M., Lallement, R., Crifo, F., & Welsh, B. Y. 1999, A&A, 346, 785

Shull, J. M. et al. 2000, ApJ, 538, L73

Snowden, S. L., Egger, R., Finkbeiner, D. P., Freyberg, M. J., & Plucinsky, P. P. 1998, ApJ, 493, 715

Sonneborn, G., Tripp, T. M., Ferlet, R., Jenkins, E. B., Sofia, U. J., Vidal-Madjar, A., & Woṫniak, P. R. 2000, ApJ, 545, 277

Sonneborn, G., et al. 2002, ApJS, submitted

Stompor, R., et al. 2001, ApJ, 561, L7

Tenorio-Tagle, G. 2000, New Astronomy Reviews, 44, 365

Timmes, F. X., Truran, J. W., Lauroesch, J. T., & York, D. G. 1997, ApJ, 476, 464

Tytler, D., O'Meara, J. M., Suzuki, N., Lubin, D. 2000, Physics Reports, 333-334, 409

Vidal-Madjar, A. 1991, Advances in Space Research, 11, 97

Vidal-Madjar, A., et al. 1998, A&A, 338, 694

Watson, J. K., & Meyer, D. M. 1996, ApJ, 473, L127

Welsh, B. Y., Vedder, P. W., Vallerga, J. V., & Craig, N. 1991, ApJ, 381, 462

Welty, D. E., Hobbs, L. M., Lauroesch, J. T., Morton, D. C., Spitzer, L., & York, D. G. 1999, ApJS, 124, 465

Whittet, D. C. B., Gerakines, P. A., Hough, J. H., &  Shenoy, S. S. 2001, ApJ 547, 872

Wolff, B., Koester, D., Dreizler, S., & Haas, S. 1998, A&A, 329, 1045

Wood, B. E., Alexander, W. R., & Linsky, J. L. 1996, ApJ, 470, 1157

Wood, B.E., et al. 2001, ApJS, submitted

Wood, B. E., & Linsky, J. L. 1998, ApJ, 492, 788

York, D. G. 1983, ApJ, 264, 172

York, D. G., & Rogerson, J. B. 1976, ApJ, 203, 378

TABLE 1. Targets for the *FUSE* LISM Studies

| Object | Spectral Type | l (°) | b (°) | d (pc) | Reference |
| --- | --- | --- | --- | --- | --- |
| HZ 43A | DA | 54.11 | 84.16 | $68 \pm 13^{a}$ | Kruk et al. 2002 |
| Feige 110 | sdOB | 74.09 | −59.07 | $179^{+265}_{-67}{}^{b}$ | Friedman et al. 2002 |
| BD +28° 4211 | sdO | 81.87 | −19.29 | $104 \pm 18^{b}$ | Sonneborn et al. 2002 |
| G191-B2B | DA1 | 155.95 | 7.10 | $69^{+19}_{-12}{}^{b}$ | Lemoine et al. 2002 |
| WD0621-376 | DA | 245.41 | −21.43 | $78 \pm 23^{c}$ | Lehner et al. 2002 |
| WD1634-573 | DO | 329.88 | −7.02 | $37 \pm 3^{b}$ | Wood et al. 2002 |
| WD2211-495 | DA | 345.79 | −52.62 | $53 \pm 16^{c}$ | Hébrard et al. 2002 |

[a] Spectroscopic + parallax distance
[b] Hipparcos parallax distance
[c] Estimated, 30% error adopted

TABLE 2. Column Densities[a]

| | log (D I) | log (O I) | log (N I) | log (N II) | log (H I) | H I Data Source | References |
|---|---|---|---|---|---|---|---|
| HZ 43A | 13.15±0.02 | 14.49±0.04 | 13.51±0.03 | 13.6±0.1 | 17.93±0.03 | *EUVE* &GHRS | 1 |
| Feige 110 | 15.47±0.03 | 16.73±0.05 | --- | --- | $20.14^{+0.07}_{-0.10}$ | *IUE* | 2 |
| BD +28° 4211 | 14.99±0.03 | 16.22±0.05 | 15.55±0.06 | --- | 19.85±0.02 | STIS | 3 |
| G191-B2B | 13.40±0.04 | 14.86±0.04 | 13.87±0.04 | --- | 18.18±0.09 | STIS & GHRS | 4 |
| WD0621-376 | $13.85^{+0.05}_{-0.04}$ | 15.26±0.04 | $14.34^{+0.05}_{-0.04}$ | >13.8 | 18.70: | *EUVE* | 5 |
| WD1634-573 | 14.05±0.03 | 15.51±0.03 | 14.62±0.04 | 14.9±0.2 | $18.85^{+0.06}_{-0.07}$ | *EUVE* | 6 |
| WD2211-495 | 13.94±0.05 | 15.34±0.04 | 14.30±0.03 | 14.6±0.2 | 18.76: | *EUVE* | 7 |

[a] Uncertainties are 1σ. A colon (:) attached to a value indicates that the value is uncertain because no error was quoted in the literature.

REFERENCES. — (1) Kruk et al. (2002); (2) Friedman et al. (2002); (3) Sonneborn et al. (2002); (4) Lemoine et al. (2002); (5) Lehner et al. (2002); (6) Wood et al. (2002); (7) Hébrard et al. (2002).

TABLE 3: Ratios of Column Densities[a,b]

| | D I/H I (10^{-5}) | O I/H I (10^{-4}) | D I/O I (10^{-2}) | N I/H I (10^{-5}) | D I/N I | O I/N I | $\frac{N\ I}{N\ I + N\ II}$ |
|---|---|---|---|---|---|---|---|
| HZ 43A | 1.66± 0.14 | 3.6± 0.4 | 4.6± 0.5 | 3.8± 0.4 | 0.44± 0.04 | 9.6± 1.1 | ~ 0.44 |
| Feige 110 | 2.14±0.41 | 3.9±0.8 | $5.5\ ^{+0.8}_{-0.7}$ | --- | --- | --- | --- |
| BD +28° 4211 | 1.39±0.10 | 2.4±0.3 | 5.9±0.7 | 5.1±0.8 | 0.27±0.05 | 4.7±0.9 | --- |
| G191-B2B | $1.66\ ^{+0.45}_{-0.30}$ | $4.8\ ^{+1.3}_{-0.9}$ | 3.5±0.4 | $4.9\ ^{+1.3}_{-0.9}$ | 0.34±0.04 | 9.8±1.2 | |
| WD0621-376 | 1.41: | 3.6: | $3.9\ ^{+0.6}_{-0.5}$ | 4.3: | $0.33\ ^{+0.05}_{-0.04}$ | $8.3\ ^{+1.3}_{-1.0}$ | --- |
| WD1634-573 | 1.60±0.25 | 4.6±0.7 | 3.5±0.3 | 6.0±1.0 | 0.27±0.03 | 7.8±0.9 | ~ 0.34 |
| WD2211-495 | 1.51: | 3.8: | 4.0±0.6 | 3.5: | 0.44±0.06 | 11.0±1.3 | ~ 0.33 |
| Weighted Mean | 1.52±0.08 | 3.03±0.21 | 3.99±0.19 | 4.24±0.31 | 0.33±0.02 | 8.1±0.4 | --- |
| Fractional Standard Deviation | 12% | 30% | 20% | 20% | 22% | 28% | --- |
| Degrees of Freedom $\nu$ | 4 | 4 | 6 | 3 | 5 | 5 | --- |
| $\chi^2_\nu$ for Mean | 1.3 | 3.9 | 2.6 | 1.8 | 3.4 | 4.4 | --- |

[a] Uncertainties are 1σ.  A colon (:) attached to a value indicates that the value is uncertain because no error for N(H I) was quoted in the literature.
[b] see Table 2 for references.

TABLE 4: Ratios Along Other Sight lines for Comparison

| | Distance (pc) | D I/H I ($10^{-5}$) | O I/H I ($10^{-4}$) | D I/O I ($10^{-2}$) | N I/H I ($10^{-5}$) | D I/N I | O I/N I ($10^{-5}$) | Satellites | Ref |
|---|---|---|---|---|---|---|---|---|---|
| Survey of near ISM | --- | --- | 3.43±0.15 | --- | 7.5±0.4 | --- | 4.1±0.3 | *HST* | a,b |
| Fractional Standard Deviation | --- | --- | 11% | --- | 12% | --- | 12% | --- | --- |
| ε Ind | 3.4 | 1.6±0.2 | --- | --- | --- | --- | --- | *HST* | c |
| Capella | 12.5 | $1.60^{+0.07}_{-0.10}$ | --- | --- | --- | --- | --- | *HST* | d |
| HR 1099 | 36 | 1.46±0.05 | --- | --- | --- | --- | --- | *HST* | e |
| 31 Com | 80 | 1.5±0.2 | --- | --- | --- | --- | --- | *HST* | e |
| θ Car | 135 | 0.50±0.16 | --- | --- | --- | --- | --- | *Copernicus* | f |
| γ Cas | 188 | 1.12±0.25 | 4.6±1.0 | --- | --- | --- | --- | *Copernicus & HST* | g |
| λ Sco | 216 | 0.76±0.25 | --- | --- | --- | --- | --- | *Copernicus* | h |
| γ² Vel | 258 | $2.18^{+0.22}_{-0.19}$ | --- | --- | --- | --- | --- | *IMAPS & IUE* | i |
| ζ Pup | 430 | $1.42^{+0.15}_{-0.14}$ | --- | --- | --- | --- | --- | *IMAPS & IUE* | i |
| δ Ori A | 500 | $0.74^{+0.12}_{-0.09}$ | 2.8±0.4 | $2.6^{+0.5}_{-0.2}$ | 4.0±0.3 | $0.19^{+0.03}_{-0.02}$ | 7.1±0.9 | *IMAPS, IUE & HST* | j |

REFERENCES. — (a) Meyer et al. (1997); (b) Meyer et al. (1998) and Meyer et al. (2001); (c) Wood et al. (1996); (d) Linsky et al. (1995); (e) Piskunov et al. (1997); (f) Allen et al. (1992); (g) H I and D I from Ferlet et al. (1980), O I from Meyer et al. (1998) and Meyer (2001); (h) York (1983); (i) Sonneborn et al. (2000); (j) Jenkins et al. (1999), O I from Meyer et al. (1998).

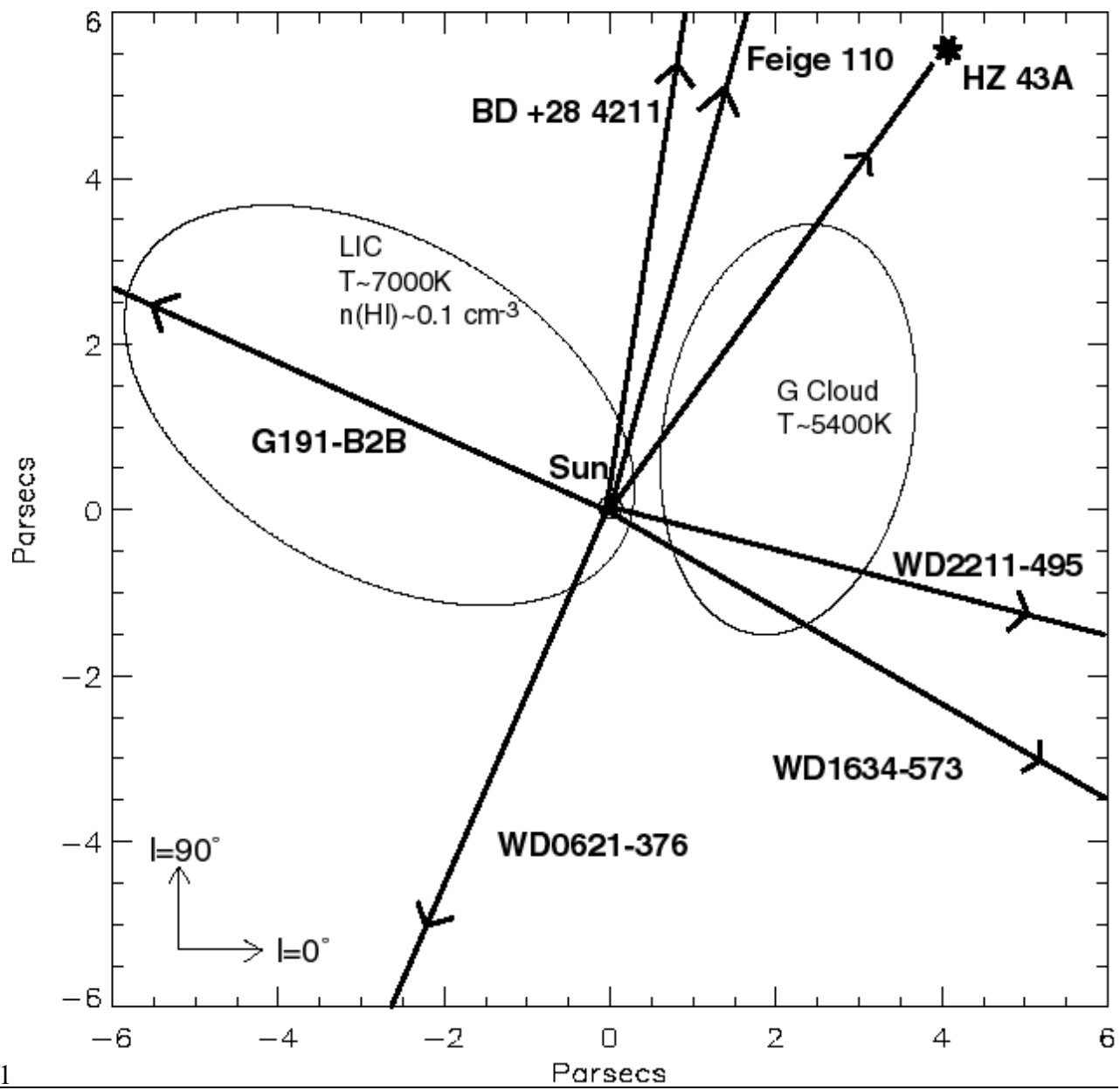

Figure 1

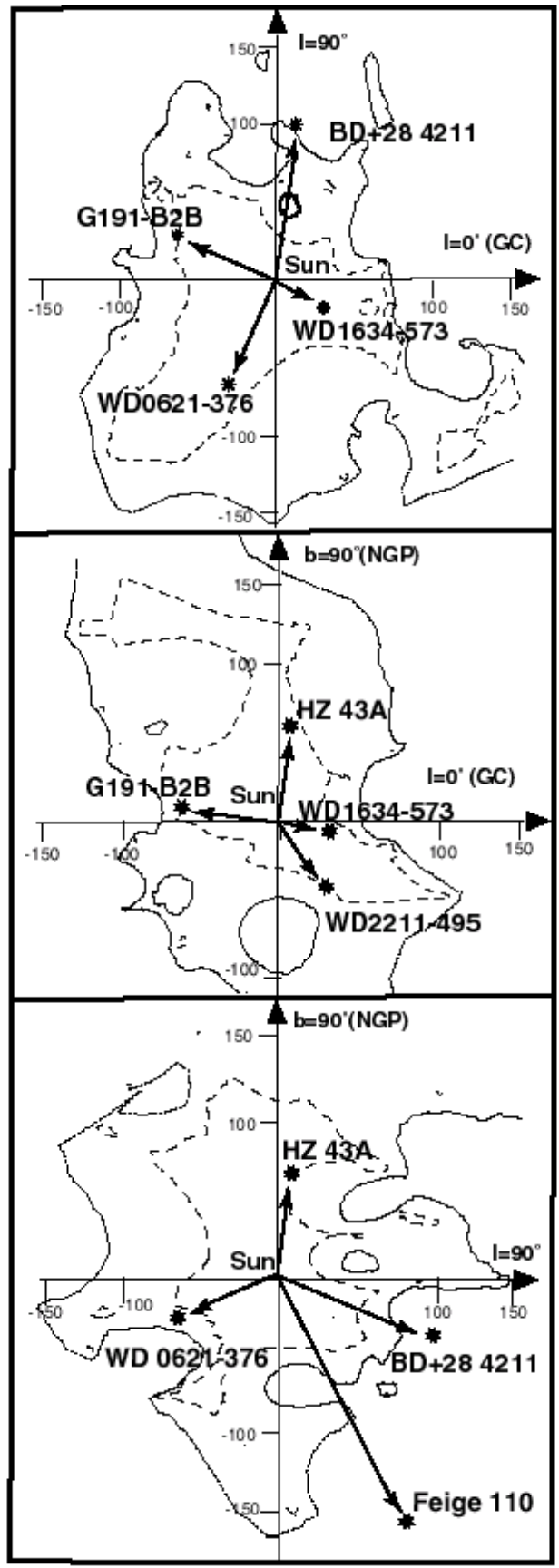

Figure 2

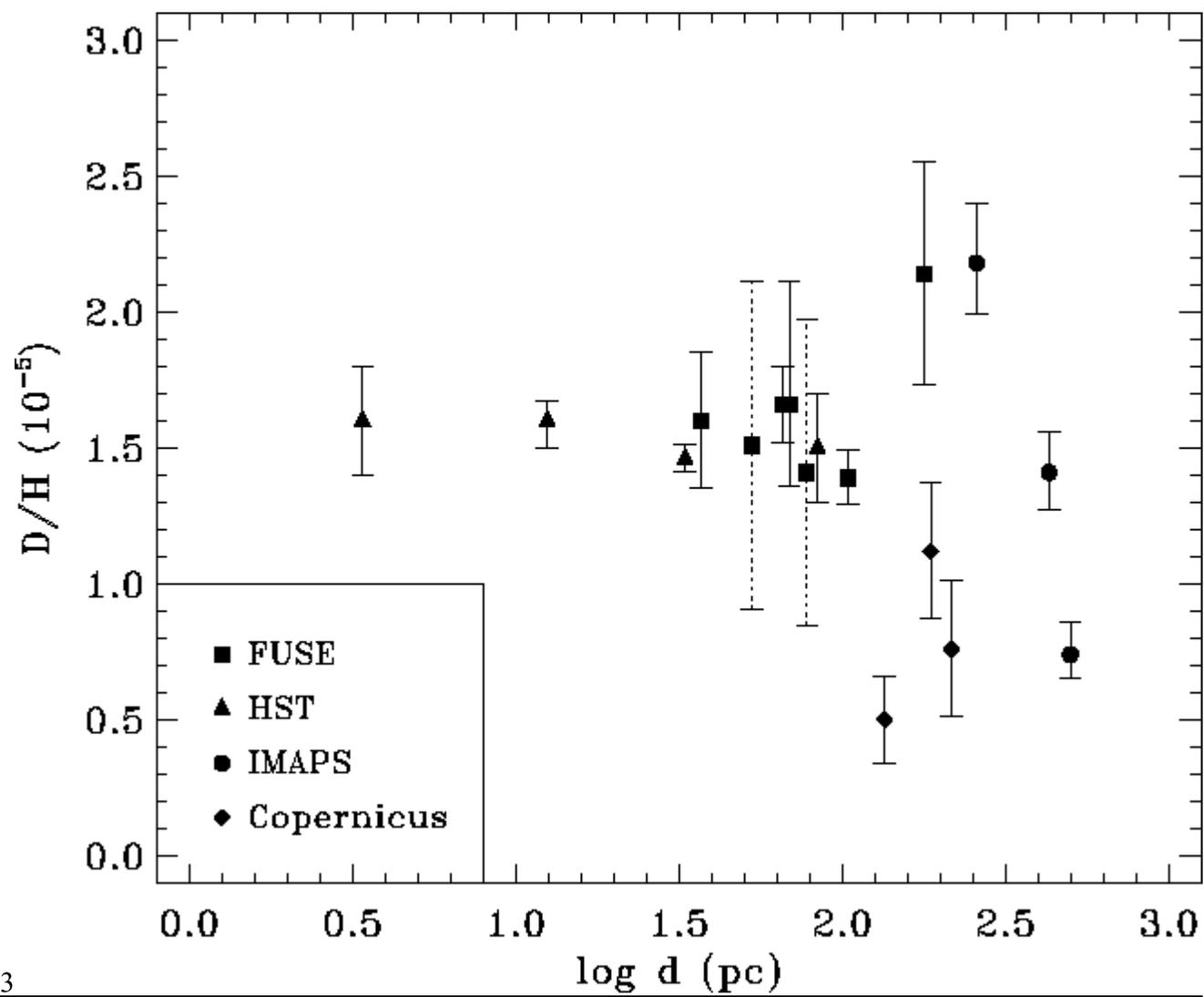

Figure 3

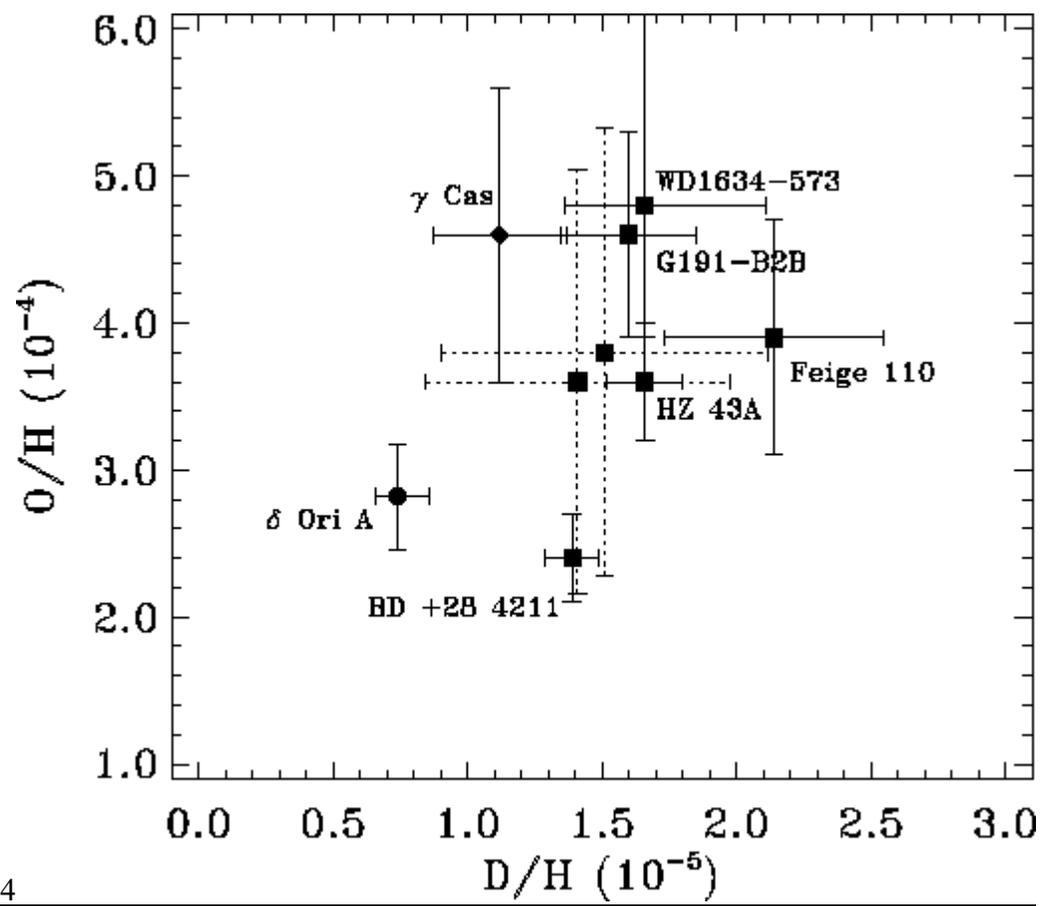

Figure 4

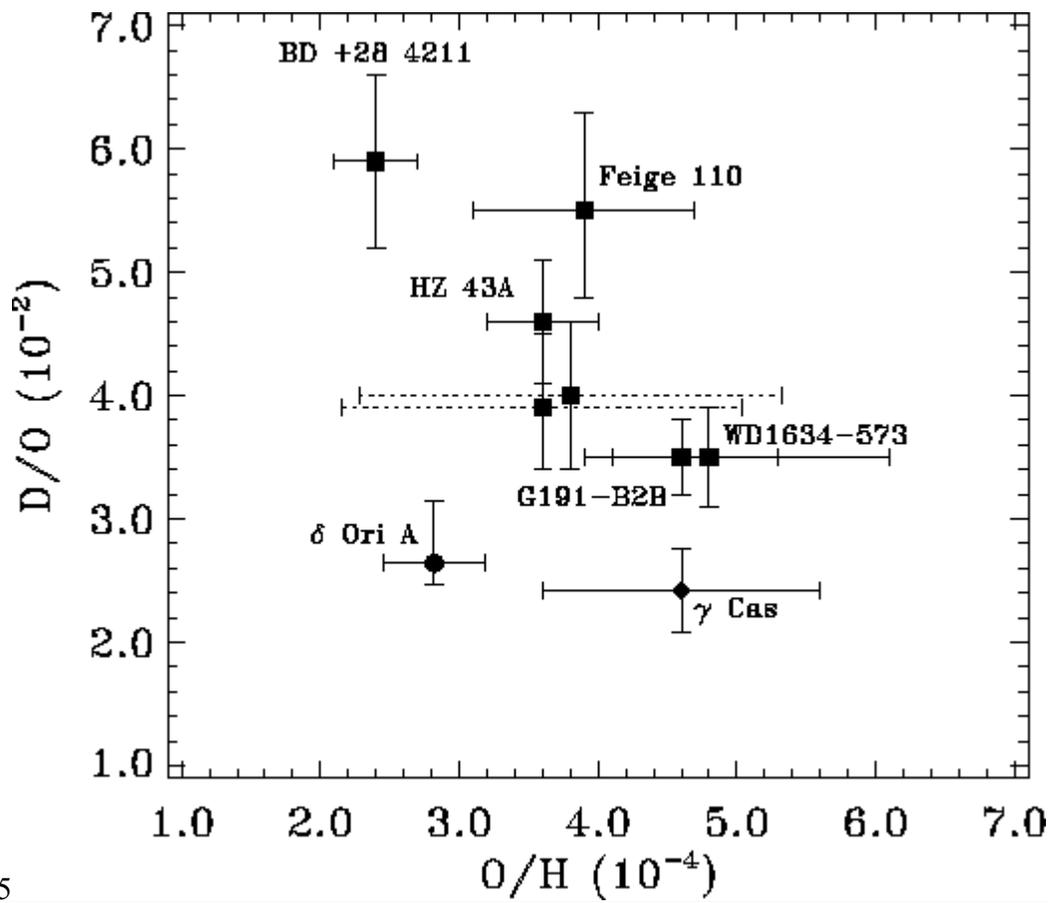

Figure 5

<u>FIGURE CAPTIONS</u>

Figure 1.  Schematic of LISM structure in the Milky Way plane near the Sun (adapted from Cha et al. 2000). Projections of the seven *FUSE* sight lines onto the Galatic plane are shown.  The HZ 43A sight line rises so steeply out of the Galactic plane that its projection does not reach the edge of the figure.

Figure 2.  Boundary of the Local Bubble and the seven *FUSE* sight lines.  Distances are in parsecs.  The 5 mÅ (dashed line) and 20 mÅ (solid line) Na I equivalent width contours are from Sfeir et al. (1999).  20 mÅ corresponds to log[N(H I)]~19.3.  The top panel is a projection on the Galactic plane.  The middle panel is the meridian plane containing the line to the Galactic Center, and the bottom panel is the meridian plane perpendicular to the line to the Galactic Center.  The projected distance to a star is plotted in planes near the sight line.  The 20 mÅ contour is dotted in regions where it is uncertain.

Figure 3.  D I/H I versus distance for the seven *FUSE* sight lines (squares) listed in Table 3.  The two data points with N(H I) determinations from *EUVE*  with large (±40%) uncertainties have dashed error bars.  Also shown are three high-confidence determinations of D / H from IMAPS (circles), 4 determinations from *HST* (triangles), and three determinations from *Copernicus* (diamonds).  Distances are from the Hipparcos catalog when available.  All errors are 1σ estimates.  See the text for references and details.

Figure 4.  O I/H I versus D I/H I for the seven *FUSE* sight lines, δ Ori A (IMAPS), and γ Cas (*Copernicus*). The two *FUSE* data points with N(H I) determinations from *EUVE* with large uncertainties have dashed error bars.  Visual inspection shows no evidence for an inverse relationship between O I/H I and D I/H I, which is contrary to expectations if D is destroyed as O is produced inside stars.

Figure 5.  D I/O I versus metallicity (O I/H I) for the seven sight lines for which accurate determinations of both N(D I) and N(O I) are available.  There is no evidence for a dependence of D I/O I on O I/H I over the limited metallicity range sampled by these sight lines.